\documentclass[preprint]{elsarticle}

\usepackage{hyperref}
\journal{Journal of Physics and Chemistry of Solids}

\makeatletter
\def\@author#1{\g@addto@macro\elsauthors{\normalsize%
    \def\baselinestretch{1}%
    \upshape\authorsep#1\unskip\textsuperscript{%
      \ifx\@fnmark\@empty\else\unskip\sep\@fnmark\let\sep=,\fi
      \ifx\@corref\@empty\else\unskip\sep\@corref\let\sep=,\fi
      }%
    \def\authorsep{\unskip,\space}%
    \global\let\@fnmark\@empty
    \global\let\@corref\@empty  
    \global\let\sep\@empty}%
    \@eadauthor={#1}
}
\makeatother

\begin{document}

\begin{frontmatter}

\title{$^{57}$Fe M\"ossbauer study of stoichiometric iron based superconductor CaKFe$_4$As$_4$: a comparison to KFe$_2$As$_2$ and CaFe$_2$As$_2$}

\author{Sergey L. Bud'ko\corref{cor1}}
\cortext[cor1]{Corresponding author} 
\ead{budko@ameslab.gov}

\author{Tai Kong\fnref{fn1}}
\fntext[fn1]{currently at Department of Chemistry, Princeton University}

\author{William R. Meier}
\address{Ames Laboratory US DOE and Department of Physics and Astronomy, Iowa State University, Ames, IA 50011, USA}

\author{Xiaoming Ma}
\address{Beijing National Laboratory for Condensed Matter Physics and Institute of Physics, Chinese Academy of Sciences, Beijing 100190, China.}
\address{Department of Physics, South University of Science and Technology of China, Shenzhen, Guangdong 518055, China}

\author{Paul C. Canfield}

\address{Ames Laboratory US DOE and Department of Physics and Astronomy, Iowa State University, Ames, IA 50011, USA}

\begin{abstract}

 $^{57}$Fe M\"ossbauer spectra at different temperatures between $\sim 5$ K and $\sim 300$ K were measured on an oriented mosaic of single crystals of CaKFe$_4$As$_4$ . The data indicate that CaKFe$_4$As$_4$ is a well formed compound with narrow spectral lines,  no traces of other,  Fe - containing, secondary phases in the spectra and no static magnetic order. There is no discernible feature at the superconducting transition temperature in any of the hyperfine parameters. The temperature dependence of the quadrupole splitting approximately follows the empirical ``$T^{3/2}$ law''.

 The hyperfine parameters of CaKFe$_4$As$_4$ are compared with those for KFe$_2$As$_2$ measured in this work, and the literature data for CaFe$_2$As$_2$, and were found to be in between those for these two, ordered, 122 compounds, in agreement with the gross view of CaKFe$_4$As$_4$ as a structural analog of  KFe$_2$As$_2$  and CaFe$_2$As$_2$ that has alternating Ca - and K - layers in the structure. 

\end{abstract}


\end{frontmatter}

\section{Introduction}
The discovery of iron-based superconductors \cite{kam08a} was followed by an outpouring of theoretical and experimental studies of those and related materials \cite{can10a,joh10a,ste11a,wan12a}. Of these studies some were addressing specific details of the superconducting and the associated vortex state, whereas others were targeted comprehensive characterization of the general physical properties of iron-based superconductors and related materials. M\"ossbauer effect spectroscopy  is widely accepted as one of the most sensitive techniques in terms of energy resolution. Historically, this technique has been applied to studies of superconductors for decades,\cite{cra61a} however its sensitivity specifically to the superconducting state is ambiguous. \cite{max15a} It is quite natural that  M\"ossbauer spectroscopy was widely used for studies of iron-based superconductors that naturally contain the common M\"ossbauer nuclide, $^{57}$Fe, in the structure,\cite{now09a,koz10a,bla11a,nat13a,jas15a}:  no additional doping with $^{57}$Fe (that can alter the properties of the material) is needed, and use of partially enriched Fe was required in only few, very specific cases. This technique was very successful in addressing the evolution of magnetic order, \cite{rot09a}  structural phase transitions, \cite{bud16a} and phase purity \cite{fel09a,rya11a,kse11a} in these materials.

\begin{figure}
    \centering
    \includegraphics[angle=0,width=80mm]{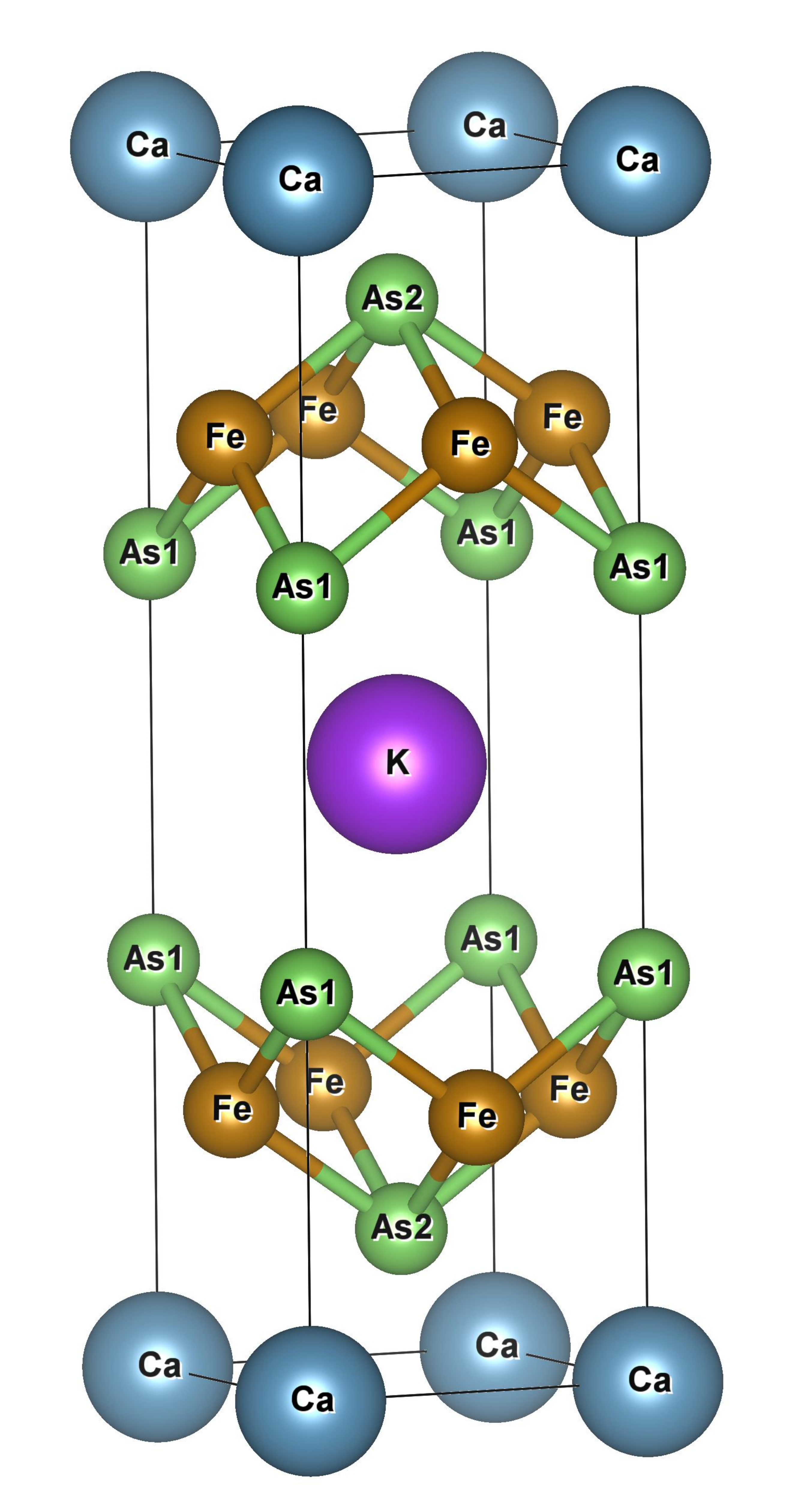}
    \caption{(Color online) Crystal structure of CaKFe$_4$As$_4$ sketched using VESTA \cite{mom11a}. }
    \label{F1}
\end{figure}

Recently, a new family of iron-based superconductors with rather high  superconducting transition temperature, $T_c \sim 31 - 36$ K,  has been discovered. \cite{iyo16a} It was found that structurally ordered CaAFe$_4$As$_4$ (1144) compounds can be formed for A = K, Rb, Cs, and the key to the formation is the difference in ionic sizes between the Ca and the A ion. This family is not a (Ca$_{1-x}$A$_x$)Fe$_2$As$_2$ solid solution, where the Ca and A ions randomly occupy a single crystallographic site. \cite{ wan13a}  Along the $c$-axis, the Ca and A ions in this family form alternating planes that are separated by the Fe-As slabs (Fig. \ref{F1}).  In essence, the CaAFe$_4$As$_4$ structure is similar to the CaFe$_2$As$_2$ structure, just with layer by layer segregation of the Ca and A ions. The ordering of the layers causes a change of the space group from $I4/mmm$ to $P4/mmm$ and the Fe site in the 1144 structure has its point symmetry lowered to orthorhombic (from the tetragonal in the 122 structure).   The 1144 structure was also found for SrAFe$_4$As$_4$ (A = Rb, Cs) \cite{iyo16a} and EuAFe$_4$As$_4$ (A = Rb,Cs). \cite{liu16a,liu16b}

We were able to grow single-crystalline, single-phase samples of CaKFe$_4$As$_4$ and measure their anisotropic thermodynamic and transport properties. \cite{mei16a} The data indicated that  CaKFe$_4$As$_4$  is an ordered stoichiometric superconductor with $T_c = 35$ K and no other phase transition for 1.8 K $\leq T \leq$ 300 K. It appeared to have properties very close to what is referred to as an optimally-doped, on a generalized phase diagram,  iron-based superconductor. Being an ordered stoichiometric compound with a high value of $T_c$ and a single crystallographic Fe site, CaKFe$_4$As$_4$ offers an exceptional opportunity to determine whether any of the hyperfine parameters exhibit an anomaly at superconducting transition. Additionally, this time with a local probe, we can evaluate the phase purity (in terms of possible Fe-containing phases) of the samples and the presence of static magnetic moment on the iron site. Furthermore, we can compare the temperature dependencies of the hyperfine parameters with those of the closely related compounds, CaFe$_2$As$_2$ and KFe$_2$As$_2$.

In this work we will present results of the $^{57}$Fe M\"ossbauer spectroscopy measurements between $\sim 5$ K and $\sim 300$ K on a mosaic of the oriented single crystals of CaKFe$_4$As$_4$ and will compare the results with similar sets of data for CaFe$_2$As$_2$ and KFe$_2$As$_2$. Whereas there are available literature data for CaFe$_2$As$_2$,  \cite{bud16a,max16a,alz11a} the published M\"ossbauer data for  KFe$_2$As$_2$ apparently is limited to three temperature points \cite{rot09a} so we have chosen to collect a comprehensive set of data for KFe$_2$As$_2$ as a part of this work.

\section{Experimental}

\begin{figure}
    \centering
    \includegraphics[angle=0,width=120mm]{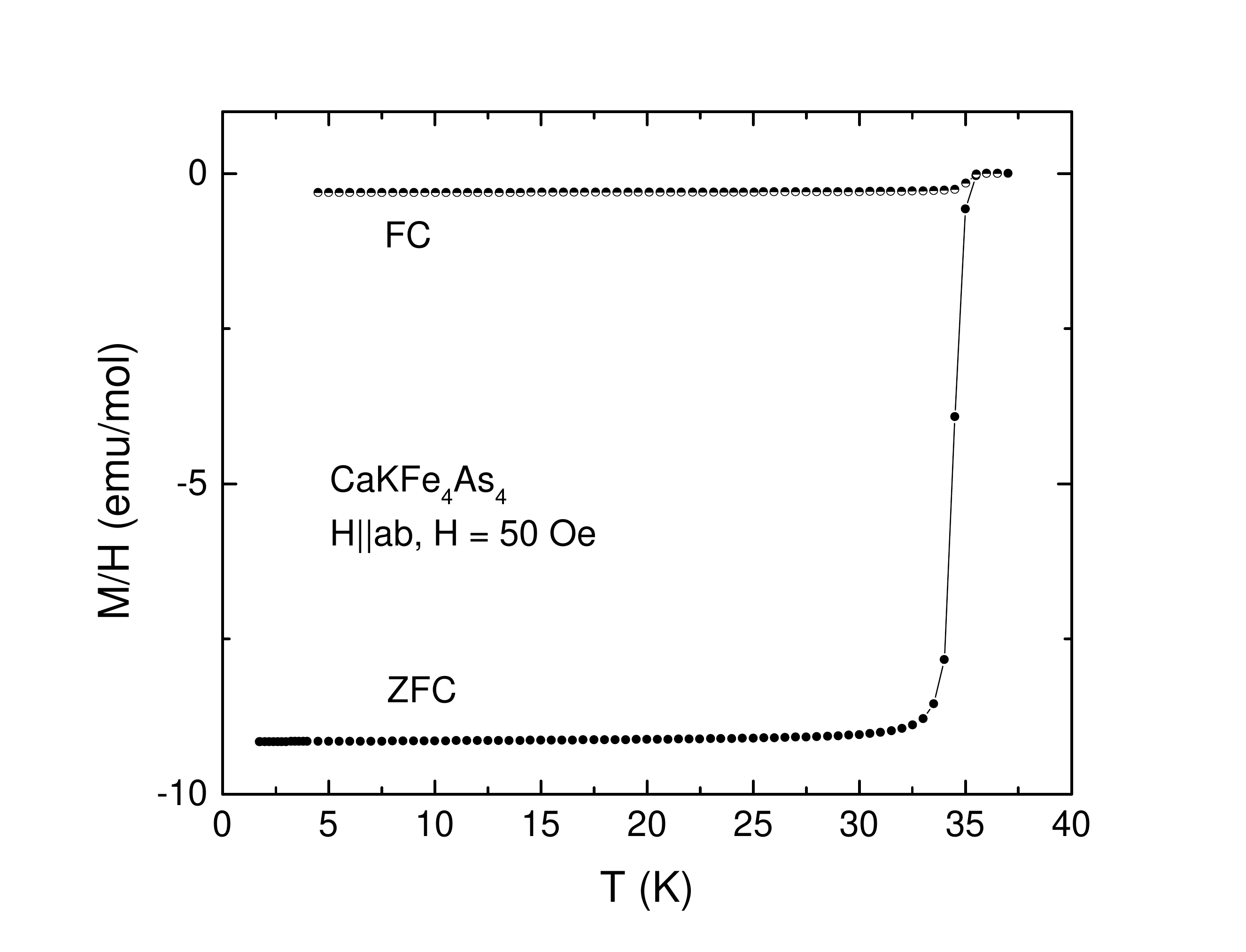}
    \caption{Low field, low temperature zero-field-cooled (ZFC) and field cooled (FC) DC susceptibility of one of the CaKFe$_4$As$_4$ crystals used in the mosaic for M\"ossbauer measurements. }
    \label{F2}
\end{figure}

CaKFe$_4$As$_4$ single crystals were grown by high temperature solution growth out of excess FeAs. The growth and basic physical properties are described in detail in Ref. \cite{mei16a}. The crystals were screened as described in Ref. \cite{mei16a} to avoid possible contaminations by CaFe$_2$As$_2$ and KFe$_2$As$_2$ minority phases. The superconducting transition in the CaKFe$_4$As$_4$  crystals used for the M\"ossbauer study was sharp with $T_c \sim 35$ K (Fig. \ref{F2}).

KFe$_2$As$_2$ single crystals were also grown using a high-temperature solution growth technique. Starting elements were packed in an alumina frit-disc crucible set \cite{can16a} with a molar ratio of K:Fe:As = 8:2:10. The crucible set together with the material were then welded in a Ta tube and sealed in a silica ampoule under a partial Ar atmosphere. A detailed drawing of such an assembly can be found in Ref \cite{kon15a}. The ampoule was slowly heated up to 920$^{\circ}$ C over $\sim 40$ hours, held at 920$^{\circ}$ C for 10 hours, quickly cooled to 850$^{\circ}$ C over 3 hours and then slowly cooled to 700$^{\circ}$ C over 3 days. At 700$^{\circ}$ C, the silica ampoule was inverted and decanted in a centrifuge. Remaining solution (primarily K-As) on the single crystals was rinsed off using ethanol. The resulting crystals had high residual resistivity ratio ($\rho (300 \textrm{K})/\rho (5 \textrm{K}) \sim 500$) and $T_c$ values consistent with other high quality KFe$_2$As$_2$ crystals. \cite{ter10a,liu13a}

M\"ossbauer spectroscopy measurements were performed using a SEE Co. conventional, constant acceleration type spectrometer in transmission geometry with a $^{57}$Co(Rh) source kept at room temperature. Both for CaKFe$_4$As$_4$ and KFe$_2$As$_2$  the absorber was prepared as a mosaic of single crystals held on a paper disk by a small amount of Apiezon N grease. The gaps between the individual crystals were kept as small as possible. The mosaic had the $c$ axis perpendicular to the disks and arbitrary in-plane orientation for each of the crystals.  The $c$ axis of the crystals in the mosaic was parallel to the M\"ossbauer $\gamma$ beam. The absorber was cooled to a desired temperature using a Janis model SHI-850-5 closed cycle refrigerator (with vibration damping). The driver velocity was calibrated using an $\alpha$-Fe foil, and all isomer shifts (IS) are quoted relative to the $\alpha$-Fe foil at room temperature. The M\"ossbauer spectra were  fitted using either the commercial software package MossWinn, \cite{kle16a} or the MossA package \cite{pre12a} with both analyses giving very similar results.

\section{Results and Discussion}

\subsection{CaKFe$_4$As$_4$}

\begin{figure}
    \centering
    \includegraphics[angle=0,width=120mm]{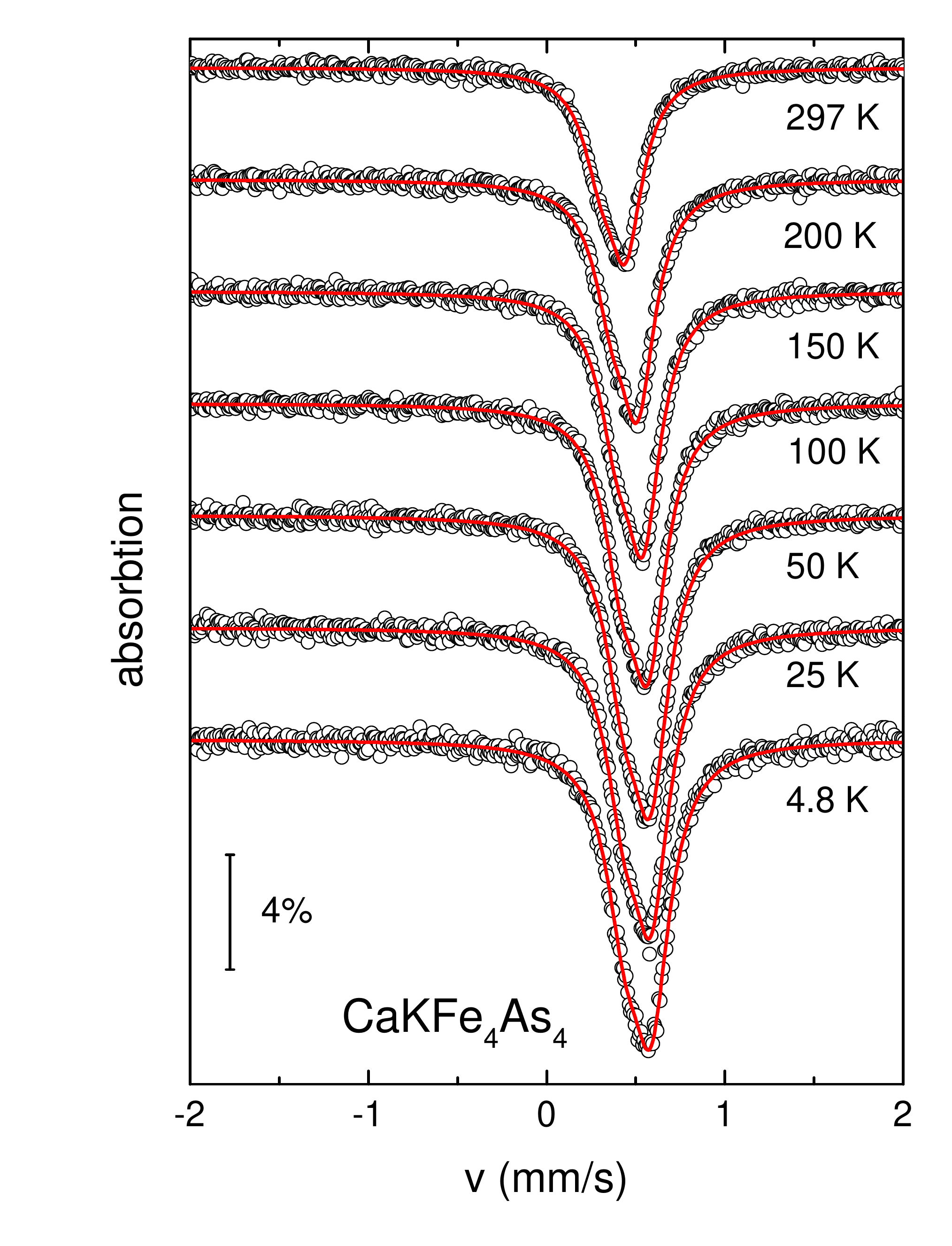}
    \caption{(Color online) $^{57}$Fe Mossbauer spectra of  CaKFe$_4$As$_4$ at selected temperatures. Symbols-data, lines-fits.}
    \label{F3}
\end{figure}

\begin{figure}
    \centering
    \includegraphics[angle=0,width=120mm]{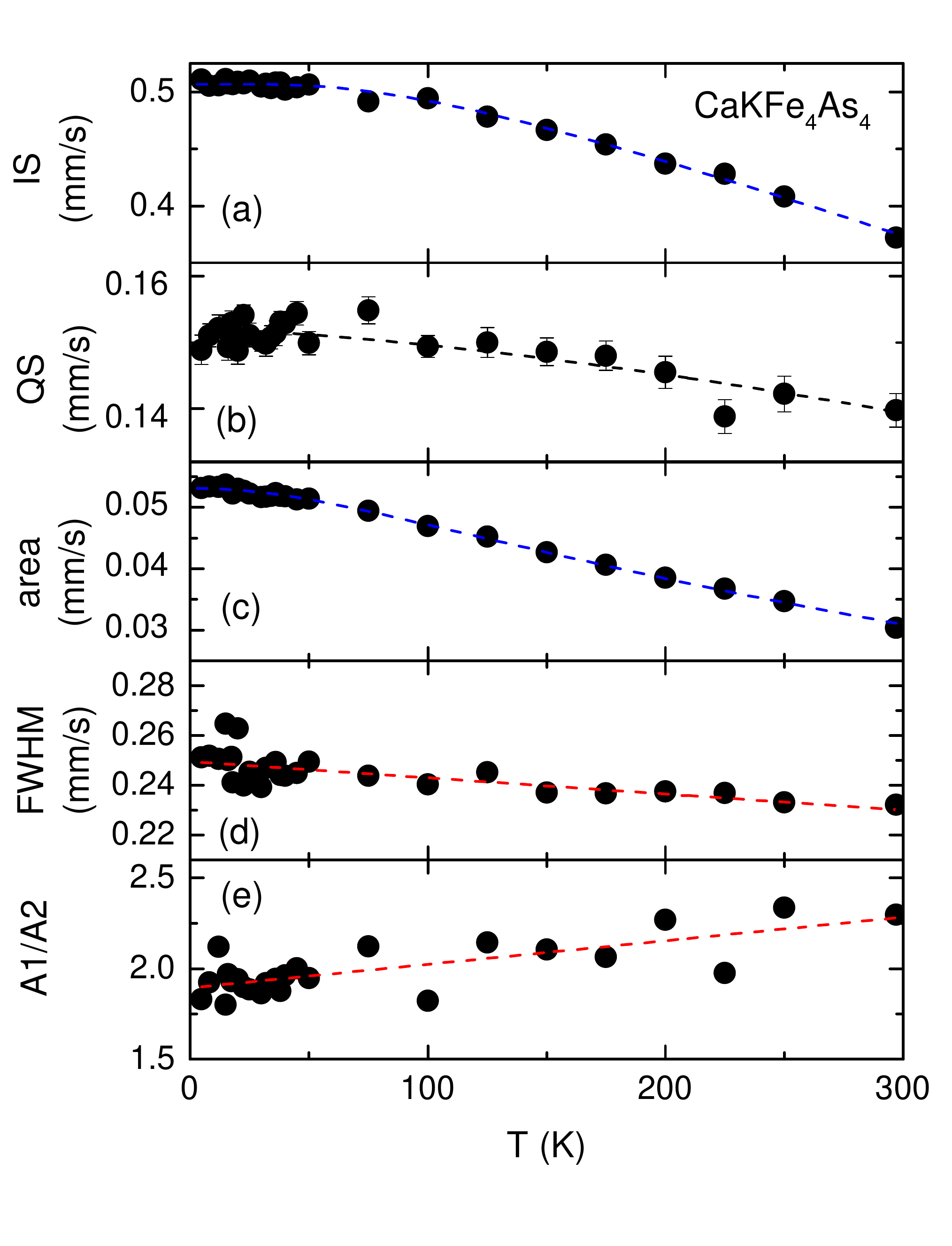}
    \caption{(Color online) Temperature dependencies of the hyperfine parameters obtained from fits of  $^{57}$Fe Mossbauer spectra of  CaKFe$_4$As$_4$ at different temperatures: (a) isomer shift, (b) quadrupole splitting, (c) area normalyzed to the baseline, (d) linewidth (full width at half maximum), and (e) intensity ratio of the doublet lines. Symbols: data, lines (a), (c) Debye fits (see text), (b)``$T^{3/2}$ law''  (see text), (d), (e) linear fits that serve as guide to the eye.}
    \label{F4}
\end{figure}

A subset of M\"ossbauer spectra for CaKFe$_4$As$_4$, taken at different temperatures, is shown in Fig. \ref{F3}. The absorption lines are asymmetric, suggesting that each spectrum is a quadrupole split doublet with rather small value of the quadrupole splitting, QS. There are no extra features observed, confirming that the samples are single phase. The results of fits to these data are shown in Fig. \ref{F4}. The linewidth of the spectra (Fig. \ref{F4}d - FWHM) varies between $\sim 0.23 - 0.26$ mm/s and is consistent with well ordered crystals. None of the hyperfine parameters has a detectable feature at $T_c = 35$ K that rises above the scattering of the data or the error bars.

Fig. \ref{F4}a presents measured  isomer shift (IS) which increases upon cooling. The isomer shift includes contributions from both the chemical shift and the second-order Doppler shift. The latter is known to increase convexly upon decreasing temperature, due to gradual depopulation of the excited phonon states, but should be constant at low temperature, because of the quantum mechanical zero-point motion. The chemical shift should not depend on temperature. The main contribution to the temperature dependence of the isomer shift then is considered to be from the second-order Doppler shift, and is usually described by the Debye model: \cite{gut11a} \\

\begin{equation}
IS(T)=IS(0)-\frac{9}{2}\frac{k_BT}{Mc}\left (\frac{T}{\Theta_D}\right )^3\int_0^ {\Theta_D/T }\frac{x^3dx}{e^x-1},
\end{equation}
\\
where $c$ is the velocity of light,  $M$ is the mass of the $^{57}$Fe nucleus, and $IS(0)$ is  the temperature-independent part. For the isomer shift data in Fig.\ref{F4}a  Debye fit yields $\Theta_D = 370 \pm 9$ K. 

The quadrupole splitting increases with decrease of temperature (Fig. \ref{F4}b). The behavior can be described reasonably well with the empirical ``$T^{3/2}$ law'' \cite{via83a}, $QS(T) = QS_0 (1-\beta T^{3/2})$, where $QS(T)$ is temperature dependent quadrupole splitting, $QS_0$ is its value at $T = 0$ K, $\beta$ is a parameter that was found \cite{via83a} to  vary between $1 \times 10^{-5}$ and $7 \times 10^{-5}$ K$^{-3/2}$. In our case the value of $\beta \approx 1.6 \times 10^{-5}$ K$^{-3/2}$ falls within the expected range.

Whereas the physics behind the ``$T^{3/2}$ law'' is not fully understood (it is considered that that it originates from thermal vibrations of the lattice \cite{nis76a}), this relation describes reasonably well the temperature dependencies of $QS$ observed in non-cubic metals. \cite{via83a,ver83a,lon99a,tam12a}

The spectral area under the doublet increases on cooling (Fig. \ref{F4}c). The temperature dependence of the spectral area can also be fitted with the Debye model \cite{gut11a}:
 \\
\begin{equation}
 f = \exp\left\{ \frac{-3E_{\gamma}^2}{k_B\Theta_DMc^2} \left[
  \frac{1}{4} + \left(\frac{T}{\Theta_D}\right)^2 
  \int_0^{\Theta_D/T}\frac{xdx}{e^x-1} \right]
 \right\},
\end{equation}
\\
where $f$ is the recoilless fraction, which is proportional to the area for a thin sample and E$_{\gamma}$ is the $\gamma$-ray energy. The estimate of the Debye temperature from the fit gives  $\Theta_D = 247 \pm 1$ K, a value that is about 125 K less than the value estimated by temperature dependence of $IS$. Although part of this discrepancy could be due to deviations from the thin absorber conditions of the measurements, it should be mentioned that  similar differences were found earlier  in studies of Lu$_2$Fe$_3$Si$_5$, FeSe$_{0.5}$Te$_{0.5}$ and $^{57}$Fe doped YBa$_2$Cu$_3$O$_{6.8}$ compounds. \cite{max15a,lin11a,che95a} This discrepancy may be explained by the fact the area reflects the average mean-square displacements, whereas $IS$ is related to the mean-square velocity of the M\"ossbauer atom. Both quantities may respond in a different way to lattice anharmonicities.

The temperature dependent linewidth of the spectra is shown in Fig. \ref{F4}d. Overall the linewidth increases by a few percent on cooling from room temperature to the base temperature. The observed spectral lines for CaKFe$_4$As$_4$ are somewhat narrower than those in CaFe$_2$As$_2$ \cite{bud16a,max16a} and KFe$_2$As$_2$ (see below) single crystal measurements, and measurably sharper than the M\"ossbauer spectra lines in the substituted Ca(Fe$_{0.965}$Co$_{0.035}$)$_2$As$_2$ \cite{boh16a} that vary in the range of 0.28 - 0.35 mm/s between room temperature and 5 K. This suggests that CaKFe$_4$As$_4$ crystals used in this work are well ordered.

In the AFe$_2$As$_2$ (A = Ba, Sr, Ca, Cs, Rb, K) compounds the point symmetry ($-4m2$) and the location of the Fe site in the crystal structure constrains the principal axis of the local electric field gradient tensor to the $c$-crystalline axis; as a result, a doublet lines intensity ratio of 3:1 is expected for the mosaic with the $c$ - axis parallel to the $\gamma$ - beam. Per contra, in the CaKFe$_4$As$_4$ the point symmetry ($2mm$) of the Fe site formally does not impose such constrain  \cite{iyo16a} and some deviation from the 3 : 1 ratio is expected. This said, the Fe - As1 and Fe - As2 bond lengths as well as As1 -  Fe -As1 and As2 - Fe - As2 bonds angles are very similar and we would not expect significant difference from the  AFe$_2$As$_2$ case.
The experimentally observed room temperature ratio is $\sim 2.3 : 1$, and it decreases to $\sim 1.9 : 1$ at the base temperature (Fig. \ref{F4}e). Very similar deviations from the 3:1 ratio were observed for measurements on CaFe$_2$As$_2$ single crystals  \cite{bud16a,alz11a,ran11a} and several possible reasons for the doublet lines intensity ratio being different from 3:1 were discussed, e.g. a “thick” absorber conditions of the measurements and some misorientation of the crystals that form the absorber mosaic. The same arguments, in addition to the different point group symmetry for Fe, are probably appropriate when considering the CaKFe$_4$As$_4$ results.

\subsection{KFe$_2$As$_2$}

\begin{figure}
    \centering
    \includegraphics[angle=0,width=120mm]{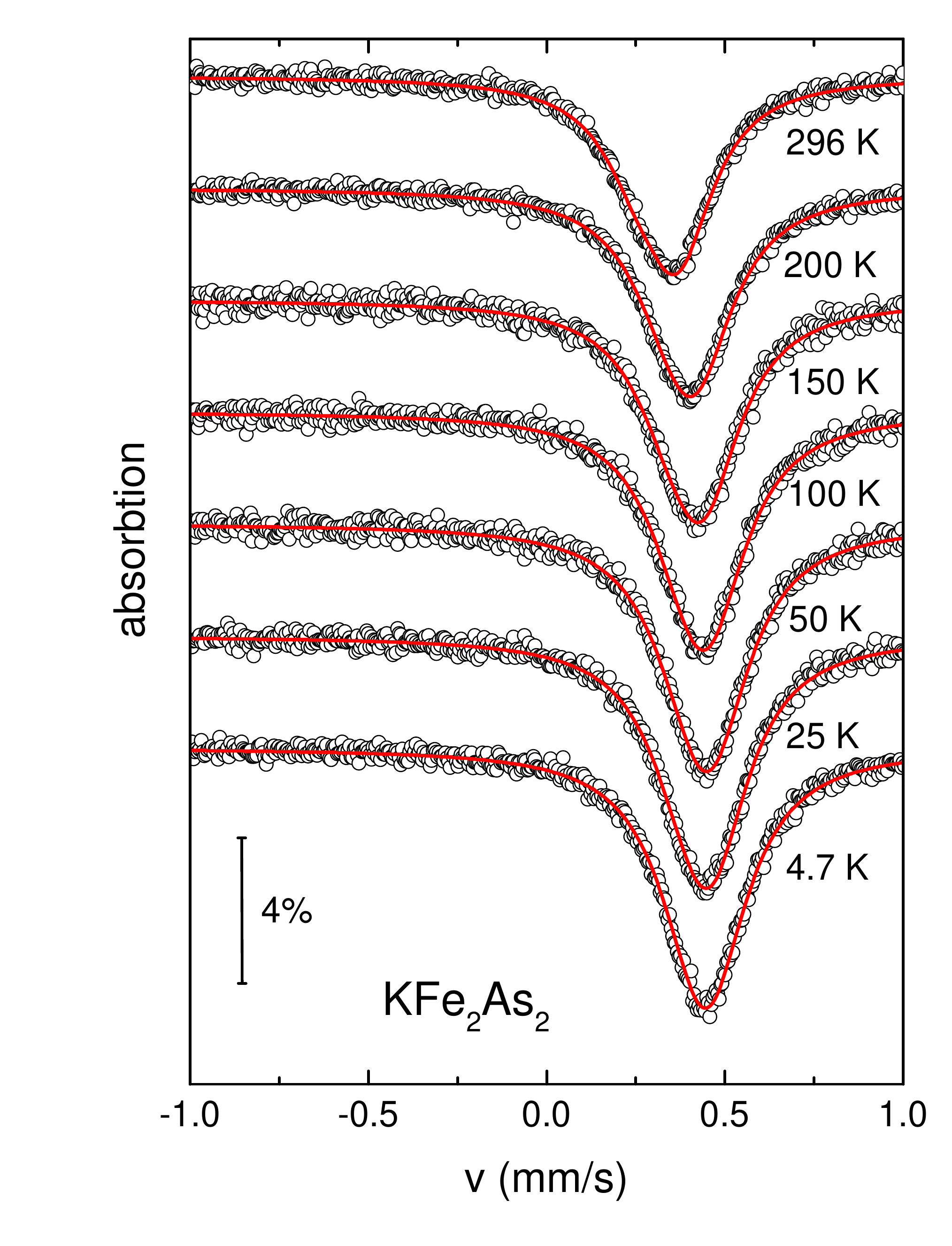}
    \caption{(Color online) $^{57}$Fe Mossbauer spectra of  KFe$_2$As$_2$ at selected temperatures. Symbols-data, lines-fits.}
    \label{F5}
\end{figure}

\begin{figure}
    \centering
    \includegraphics[angle=0,width=120mm]{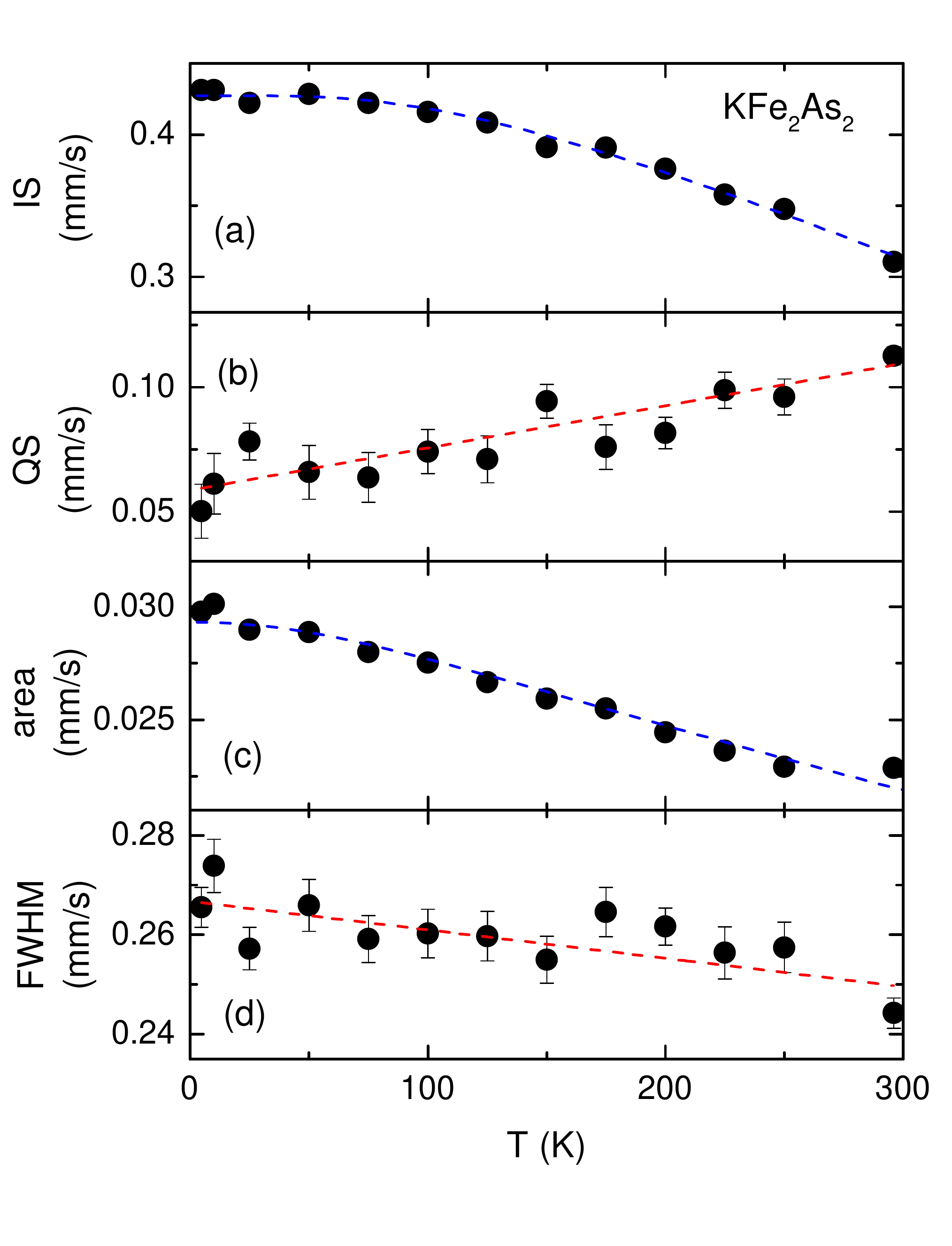}
    \caption{(Color online) Temperature dependencies of the hyperfine parameters obtained from fits of  $^{57}$Fe Mossbauer spectra of  KFe$_2$As$_2$ at different temperatures: (a) isomer shift, (b) quadrupole splitting, (c) area normalyzed to the baseline, (d) linewidth (full width at half maximum). Symbols: data, lines (a), (c) Debye fits (see text), (b) and (d) linear fits that serve as guide to the eye.}
    \label{F6}
\end{figure}

Fig. \ref{F5} shows a subset of M\"ossbauer spectra of KFe$_2$As$_2$ taken at different temperatures. The asymmetry is even less pronounced than in the CaKFe$_4$As$_4$ spectra above. Still, good fit of the data can be obtained by using a doublet with small quadrupole splitting. For KFe$_2$As$_2$, similarly to the CaKFe$_4$As$_4$, the principal axis of the local electric field gradient tensor should be parallel to the $c$-crystalline axis and the doublet lines intensity ratio of 3:1 is expected. If the fits are performed with this ratio left as a free parameter, within the error bars the expected $A1/A2 = 3$ is obtained for all temperatures. To reduce the uncertainty in particular, in the small values of  quadrupole splitting, we repeated the fits with fixed the  $A1/A2 = 3$ parameter.  The results are shown in Fig. \ref{F6}. In comparison with the data obtained on powders at three different temperatures, \cite{rot09a}, our linewidth is smaller and values of the isomer shift are bigger, although the overall changes, $\Delta IS$, from room temperature to the base temperatures are very similar.

The linewidth and its temperature dependence  (Fig. \ref{F6}d) are similar to those observed for CaKFe$_4$As$_4$. The temperature dependent isomer shift and spectral area are well fit using the Debye model, as described above  (Fig. \ref{F6}a,c). These fits yield the values of $\Theta_D$ of $474 \pm 20$ K (from $IS(T)$) and $325 \pm 7$ K (from temperature dependent spectral area). In comparison with CaKFe$_4$As$_4$, the Debye temperatures are higher in KFe$_2$As$_2$, suggesting that the lattice is stiffer. 

The values of quadrupole splitting (Fig. \ref{F6}b)  for KFe$_2$As$_2$ are significantly smaller than those for CaKFe$_4$As$_4$, moreover $QS$ decreases with decrease of temperature, as opposed to increase following the ``$T^{3/2}$ law'' in CaKFe$_4$As$_4$. Although the theoretical foundations of the empirical  ``$T^{3/2}$ law''  are not well understood and different (constant, vs $T^{3/2}$) $QS(T)$ behavior has been observed for related  (Ce$_{0.35}$FeCo$_3$Sb$_{12}$  vs Ce$_{0.98}$Fe$_4$Sb$_{12}$)  materials,\cite{lon99a} this observation in iron-arsenides calls for further studies.

\subsection{Comparison of CaKFe$_4$As$_4$, KFe$_2$As$_2$, and CaFe$_2$As$_2$}

Since, naively speaking, the CaKFe$_4$As$_4$ structure can be viewed as being costructed from the alternating slabs of CaFe$_2$As$_2$ and KFe$_2$As$_2$ structures, it would be of use to compare the $^{57}$Fe  hyperfine parameters of these three compounds. Whereas CaKFe$_4$As$_4$ and KFe$_2$As$_2$ do not exhibit static magnetic order or a structural transition below room temperature, CaFe$_2$As$_2$ is known to be more complex. CaFe$_2$As$_2$ grown out of Sn flux \cite{nin08a} exhibits concomitant structural (high temperature tetragonal to low temperature orthorhombic) and magnetic (paramagnetic to low temperature antiferromagnetic) transitions at $\approx 173$ K. \cite{nin08a,gol08a}. In the following we will refer to this sample as CaFe$_2$As$_2$ - AFM and use the hyperfine parameters from the single crystal work, Ref. \cite{max16a}. In the CaFe$_2$As$_2$ sample grown out of FeAs flux, by judicious choice of annealing / quenching conditions \cite{ran11a} we can stabilize low temperature ambient pressure collapsed tetragonal (cT) phase with the structural transition at $\approx 90$ K. This sample will be referred in the following as CaFe$_2$As$_2$ - cT, and the hyperfine parameters from the ref. \cite{bud16a} will be used.

\begin{figure}
    \centering
    \includegraphics[angle=0,width=120mm]{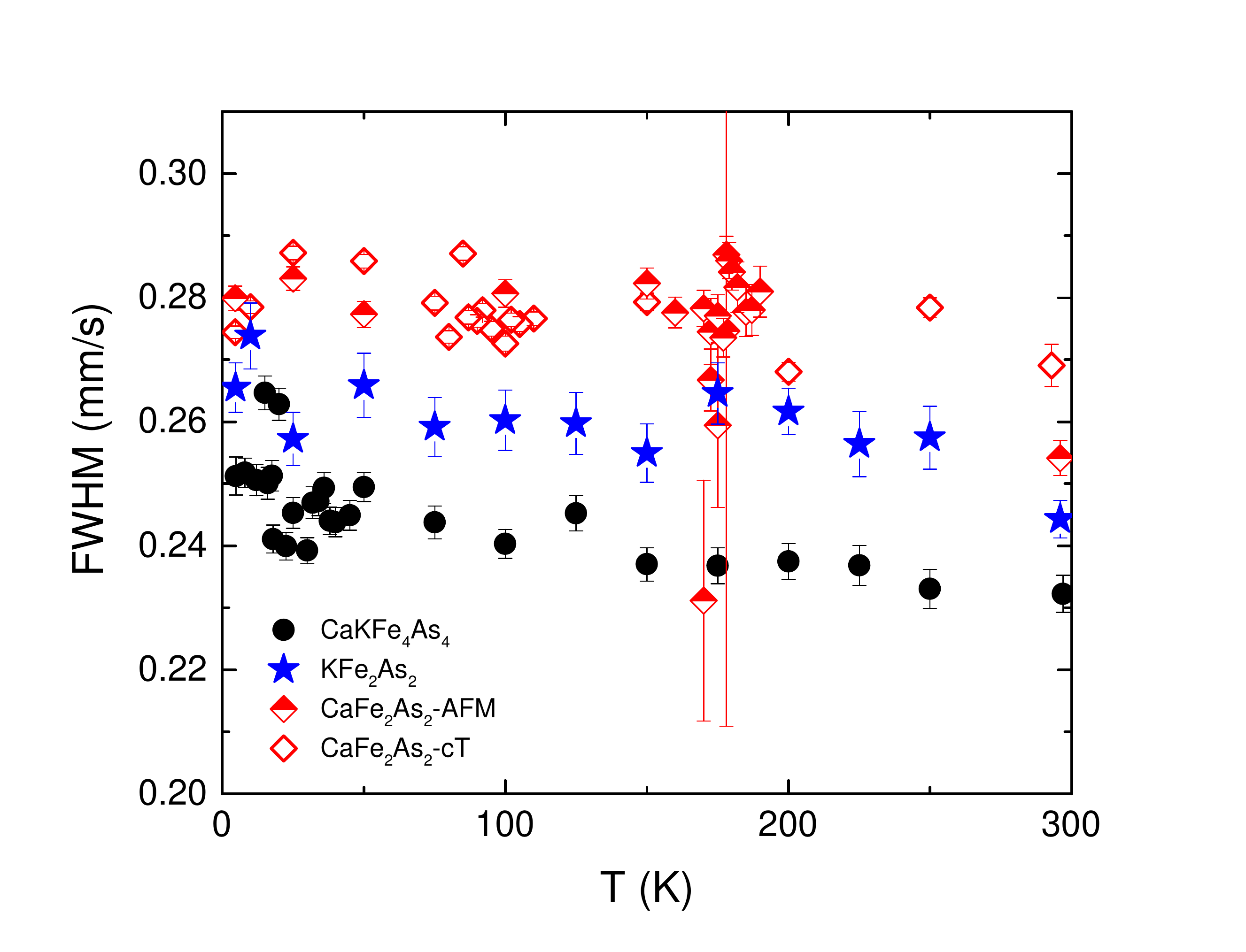}
    \caption{(Color online) Temperature dependencies of the  linewidth (full width at half maximum) obtained from fits of  $^{57}$Fe Mossbauer spectra of  CaKFe$_4$As$_4$, KFe$_2$As$_2$ (this work) and CaFe$_2$As$_2$ \cite{bud16a,max16a} at different temperatures.}
    \label{FWHM}
\end{figure}

\begin{figure}
    \centering
    \includegraphics[angle=0,width=120mm]{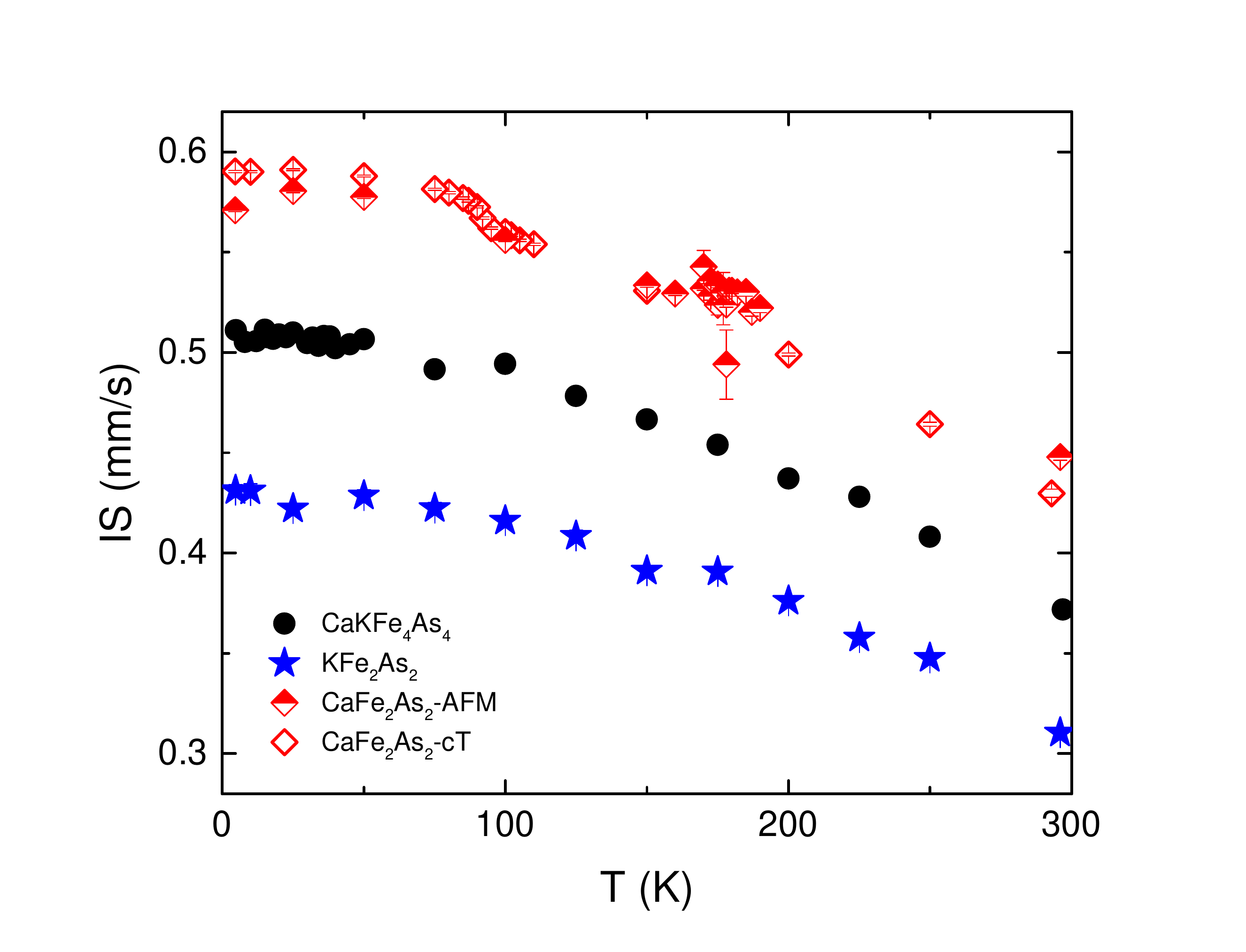}
    \caption{(Color online) Temperature dependencies of the  isomer shift obtained from fits of  $^{57}$Fe Mossbauer spectra of  CaKFe$_4$As$_4$, KFe$_2$As$_2$ (this work) and CaFe$_2$As$_2$ \cite{bud16a,max16a} at different temperatures.}
    \label{IS}
\end{figure}

\begin{figure}
    \centering
    \includegraphics[angle=0,width=120mm]{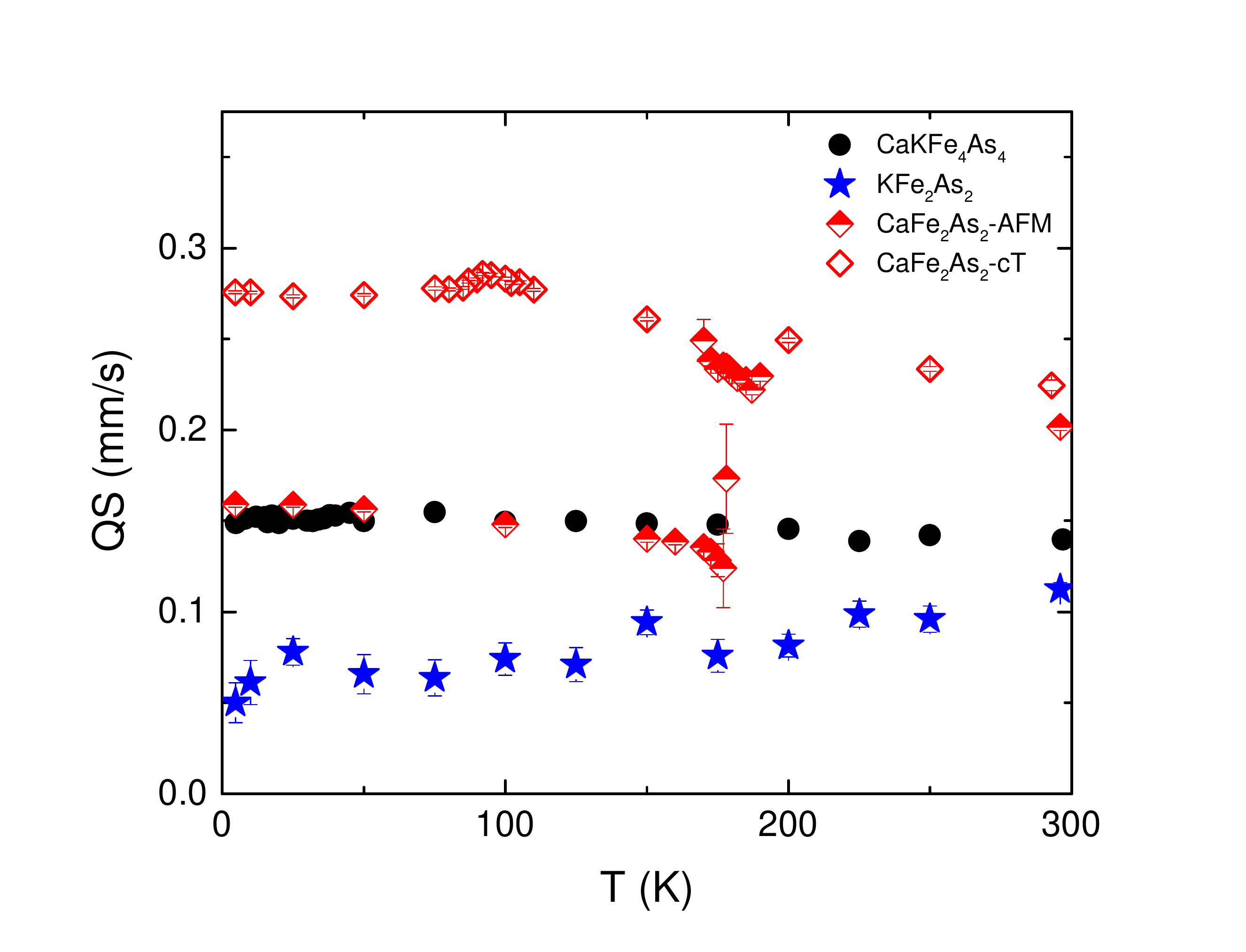}
    \caption{(Color online) Temperature dependencies of the quadrupole splitting obtained from fits of  $^{57}$Fe Mossbauer spectra of  CaKFe$_4$As$_4$, KFe$_2$As$_2$ (this work) and CaFe$_2$As$_2$ \cite{bud16a,max16a} at different temperatures. Note that for CaFe$_2$As$_2$ - AFM the absolute values of the quadrupole splitting are plotted in the magnetically ordered state.}
    \label{QS}
\end{figure}

The hyperfine parameters for the  CaKFe$_4$As$_4$, KFe$_2$As$_2$, and CaFe$_2$As$_2$ compounds at room temperature and the base temperature are summarized in the Table \ref{T1}. The temperature dependencies are presented in the plots below.  The spectral  linewidths of these compounds are very similar (Fig. \ref{FWHM}). The smallest one is observed for CaKFe$_4$As$_4$, possibly pointing out to very well formed crystals. The isomer shift in CaKFe$_4$As$_4$ (Fig. \ref{IS}). has values in between those for KFe$_2$As$_2$ and CaFe$_2$As$_2$. Note that the $IS(T)$ for  CaFe$_2$As$_2$ - cT has a small but distinct feature associated with the cT transition. Similarly, the quadrupole splitting of  CaKFe$_4$As$_4$ (Fig. \ref{QS}) has values in between the values for two other compounds. Both, CaFe$_2$As$_2$ - cT and CaFe$_2$As$_2$ - AFM have clear features associated with the cT and the structural / AFM transitions, respectively. It is curious (although it might be a mere coincidence) that $QS(T)$ of CaKFe$_4$As$_4$ and the absolute values of $QS(T)$ of CaFe$_2$As$_2$ - AFM below the structural / AFM transition are laying basically on top of each other. As for the overall temperature behavior, it appears that only for KFe$_2$As$_2$ $QS(T)$ decreases with decrease of temperature. It would be of interest to see if different temperature dependences are observed in quadrupole frequencies when measured in these materials by $^{75}$As nuclear magnetic resonance. 
As for normalized (at the respective base temperatures) spectral areas (Fig. \ref{F10}), again, CaFe$_2$As$_2$ - cT and CaFe$_2$As$_2$ - AFM have anomalies at the corresponding transitions. The data points for CaKFe$_4$As$_4$, CaFe$_2$As$_2$ - cT and CaFe$_2$As$_2$ - AFM (below the structural / AFM transition) are, grossly speaking, following the same temperature dependence. The data for KFe$_2$As$_2$ are somewhat distinct, pointing either to stiffer phonon spectrum or some additional contribution to the temperature dependence of the spectral area in this compound.

\section{Summary}

The measurements of $^{57}$Fe M\"ossbauer spectra on oriented mosaics of single crystals of CaKFe$_4$As$_4$ and  KFe$_2$As$_2$ were performed and the results were compared with the literature data for CaFe$_2$As$_2$.

CaKFe$_4$As$_4$ can be characterized as a well formed compound with narrow spectral lines and  no traces of other,  Fe - containing, secondary phases in the spectra. There is no feature in hyperfine parameters at $T_c$ and no indication of static magnetic order. The values of the $^{57}$Fe hyperfine parameters of CaKFe$_4$As$_4$ are in between those for KFe$_2$As$_2$  and CaFe$_2$As$_2$, in agreement with the gross view of CaKFe$_4$As$_4$ as a structural analog of  KFe$_2$As$_2$  and CaFe$_2$As$_2$ with  alternating Ca - and K - layers in the structure. The $QS(T)$ generally follows the empirical ``$T^{3/2}$ law''. Debye fits of the temperature dependencies of the isomer shift and the spectral area yield the Debye temperatures of $\sim 370$ K and $\sim 247$ K respectively. 

KFe$_2$As$_2$ has smaller quadrupole splitting and isomer shift in comparison with  CaKFe$_4$As$_4$ and CaFe$_2$As$_2$. Its $QS$ decreases slightly on cooling that differs from the generic behavior observed in many non-cubic metals. The Debye temperatures evaluated from the temperature dependent $IS$ and spectral area are $\sim 474$ K and $\sim 325$ K respectively, these values being $\sim 100$ K higher than those for CaKFe$_4$As$_4$.

\section*{Acknowledgments}
This work was supported by the U.S. Department of Energy, Office of Basic Energy Science, Division of Materials Sciences and Engineering. The research was performed at the Ames Laboratory. Ames Laboratory is operated for the U.S. Department of Energy by Iowa State University under Contract No. DE-AC02-07CH11358. In addition,  W. R. M. was supported by
the Gordon and Betty Moore Foundations EPiQS Initiative through Grant GBMF4411.

\clearpage

\begin{table}
\caption{Hyperfine parameters of CaKFe$_4$As$_4$, KFe$_2$As$_2$, and CaFe$_2$As$_2$ at room temperature and base temperature} \label{T1}

\begin{tabular}{ l r c c c l }
  \hline
  \hline			
  sample & $T$ (K) & $IS$ (mm/s) & $QS$ (mm/s) & FWHM (mm/s) & Ref. \\
  \hline
  \hline

  CaKFe$_4$As$_4$ & 297 & 0.372(2) & 0.140(3) & 0.232(3) & this work \\
   & 4.8 & 0.511(2) & 0.149(2) & 0.251(3) & this work\\		
\hline

KFe$_2$As$_2$ & 296 & 0.311(1) & 0.113(4) & 0.244(3) & this work \\				
   & 4.7 & 0.431(3) & 0.05(1) & 0.265(4) & this work\\
\hline

CaFe$_2$As$_2$ - AFM & 296 & 0.448(2) & 0.202(2) & 0.254(3) & \cite{max16a} \\
   & 4.6 & 0.5708(7) & -0.159(1) & 0.280(2) & \cite{max16a}\\
\hline

CaFe$_2$As$_2$ - cT & 293 & 0.430(2) & 0.225(3) & 0.269(3) & \cite{bud16a} \\
   & 4.6 & 0.5902(5) & 0.2758(7) & 0.274(1) & \cite{bud16a}\\

  \hline  
\end{tabular}

\end{table}

\clearpage

\clearpage

\clearpage

\clearpage

\clearpage

\clearpage

\clearpage

\clearpage

\clearpage

\clearpage

\begin{figure}
    \centering
    \includegraphics[angle=0,width=120mm]{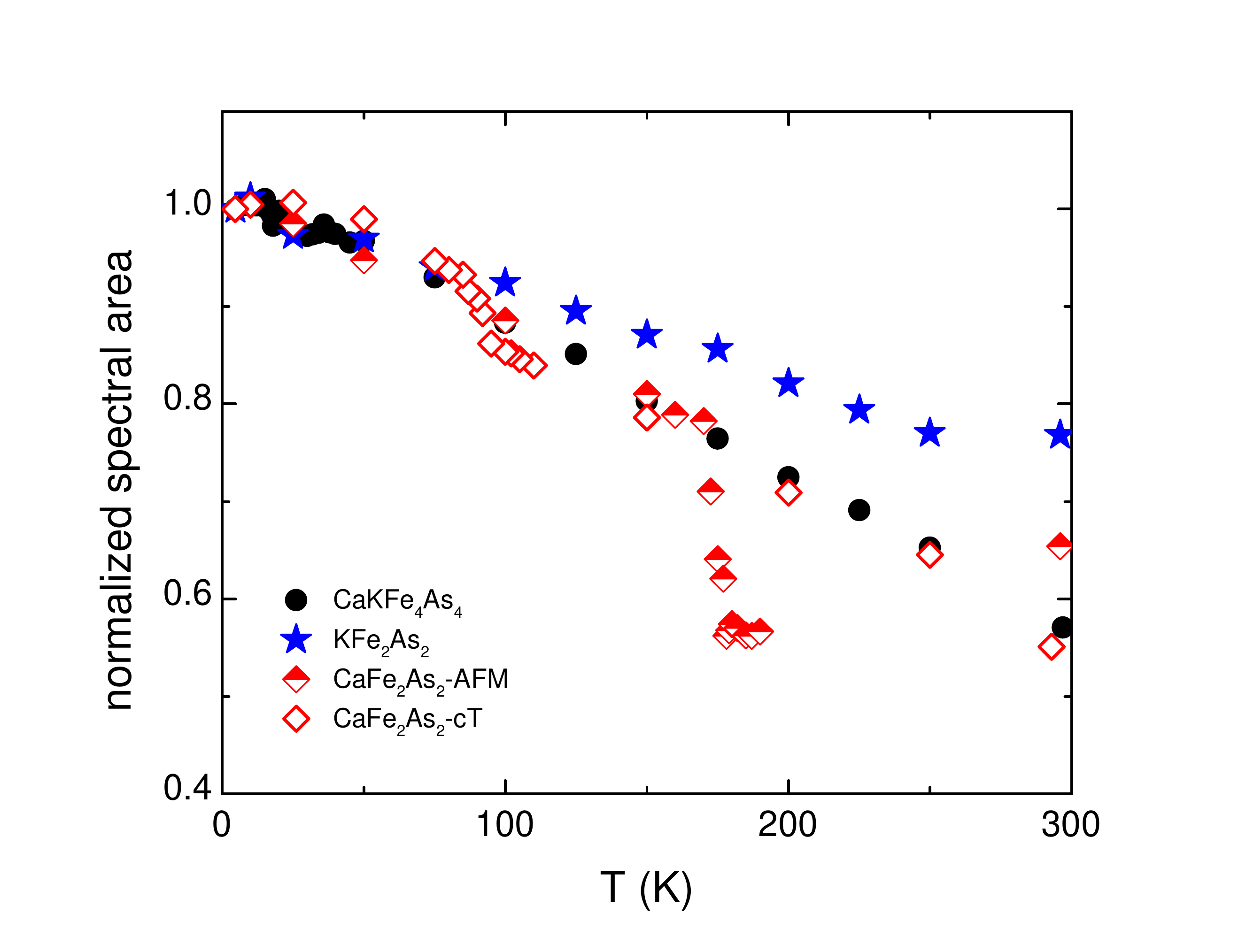}
    \caption{(Color online) Temperature dependencies of the  spectral area normalyzed to the corresponding value at the base temperature (4.6 K - 4.8 K)  obtained from fits of  $^{57}$Fe Mossbauer spectra of  CaKFe$_4$As$_4$, KFe$_2$As$_2$ (this work) and CaFe$_2$As$_2$ \cite{bud16a,max16a} at different temperatures.}
    \label{Narea}
\end{figure}


\section*{References}

\begin{thebibliography}{1}

\bibitem{kam08a}
 Y. Kamihara, T. Watanabe, M. Hirano, H. Hosono, J. Am. Chem. Soc. 130 (2008) 3296.

\bibitem{can10a}
P. C.   Canfield and S. L.   Bud'ko, Annu. Rev. Condens. Matter Phys.   1  (2010) 27.

\bibitem{joh10a}
D. C.   Johnston, Adv. Phys.   59  (2010) 803.

\bibitem{ste11a}
G. R.   Stewart, Rev. Mod. Phys.   83 (2011) 1589.

\bibitem{wan12a}
N.-L.   Wang, H.   Hosono and P.-C.   Dai (eds.), Iron-based Superconductors. Materials, Properties and Mechanisms (Pan Stanford Publishing , Boca Raton, FL , 2012) .

\bibitem{cra61a}
P. P. Craig, R. D. Taylor, and D. E. Nagle, Nuovo Cim. 22 (1961) 402.

\bibitem{max15a}
Xiaoming Ma, Sheng Ran, Hua Pang, Fashen Li, Paul C. Canfield, and Sergey L. Bud'ko, J. Phys. Chem. Solids 83 (2015) 58.

\bibitem{now09a}
Israel Nowik, Israel Felner, Physica C 469 (2009) 485.

\bibitem{koz10a}
M. G. Kozin, I. L. Romashkina, Izv. Ros. Akad. Nauk, Ser. Fiz. 74 (2010) 360 [Bull. Rus. Acad. Sci.: Physics 74 (2010) 330].

\bibitem{bla11a}
A. B{\l}achowski, K. Ruebenbauer, and J. \.Zukrowski, The Annales UMCS, Sectio AAA - Physica 66 (2011) 125.

\bibitem{nat13a}
Amar Nath and Airat Khasanov, in: M\"ossbauer Spectroscopy: Applications in Chemistry, Biology, and Nanotechnology, edited by Virender K. Sharma, Gostar Klingelhofer, and Tetsuaki Nishida (John Wiley \& Sons, Inc., Hoboken, NJ, 2013), p. 535.

\bibitem{jas15a}
A. K. Jasek, K. Kom\c edera, A. B{\l}achowski, K. Ruebenbauer, J. \.Zukrowski, Z. Bukowski, and J. Karpinski, Philos. Mag. 95 (2015) 493.

\bibitem{rot09a}
Marianne Rotter,  Marcus Tegel, Inga Schellenberg, Falko M. Schappacher, Rainer P\"ottgen, Joachim Deisenhofer, Axel G\"unther, Florian Schrettle, Alois Loidl, and Dirk Johrendt, New J. Phys. 11 (2009) 125014.

\bibitem{bud16a}
Sergey L. Bud'ko, Xiaoming Ma, Milan Tomi\'c, Sheng Ran, Roser Valent\'i, and Paul C. Canfield, Phys. Rev. B 93 (2016) 024516.

\bibitem{fel09a}
Israel Felner, Israel Nowik, Bing Lv, Joshua H. Tapp, Zhongjia Tang, and Arnold M. Guloy, Hyperfine Int. 191 (2009) 61.

\bibitem{rya11a}
D. H. Ryan, W. N. Rowan-Weetaluktuk, J. M. Cadogan, R. Hu, W. E. Straszheim, S. L. Bud'ko, and P. C. Canfield, Phys. Rev. B 83 (2011) 104526.

\bibitem{kse11a}
Vadim Ksenofontov, Gerhard Wortmann, Sergey A. Medvedev, Vladimir Tsurkan, Joachim Deisenhofer, Alois Loidl, and Claudia Felser, Phys. Rev. B 84 (2011) 180508.

\bibitem{iyo16a}
Akira Iyo, Kenji Kawashima, Tatsuya Kinjo, Taichiro Nishio, Shigeyuki Ishida, Hiroshi Fujihisa, Yoshito Gotoh, Kunihiro Kihou, Hiroshi Eisaki, and Yoshiyuki Yoshida, J. Amer. Chem. Soc. 138 (2016) 3410.

\bibitem{wan13a}
D. M. Wang, X. C. Shangguan, J. B. He, L. X. Zhao, Y. J. Long, P. P. Wang, and L. Wang, J. Supercond. Nov. Magn. 26 (2013) 2121.

\bibitem{mom11a}
K. Momma and F. Izumi, J. Appl. Crystallogr., 44 (2011) 1272.

\bibitem{liu16a}
Yi Liu, Ya-Bin Liu, Zhang-Tu Tang, Hao Jiang, Zhi-Cheng Wang, Abduweli Ablimit, Wen-He Jiao, Qian Tao, Chun-Mu Feng, Zhu-An Xu, and Guang-Han Cao, Phys. Rev. B 93 (2016) 214503.

\bibitem{liu16b}
Yi Liu, Ya-Bin Liu, Qian Chen, Zhang-Tu Tang, Wen-He Jiao, Qian Tao, Zhu-An Xu, Guang-Han Cao, Science Bulletin 61 (2016) 1213.

\bibitem{mei16a}
W. R. Meier, T. Kong, U. S. Kaluarachchi, V. Taufour, N. H. Jo, G. Drachuck, A. E. B\"ohmer, S. M. Saunders, A. Sapkota, A. Kreyssig, M. A. Tanatar, R. Prozorov, A. I. Goldman, Fedor F. Balakirev, Alex Gurevich, S. L. Bud'ko, and P. C. Canfield, Phys. Rev. B 94 (2016) 064501.

\bibitem{max16a}
Xiaoming Ma,  Sheng Ran, Paul C. Canfield, Sergey L. Bud'ko, J. Alloys Compd 657 (2016) 379.

\bibitem{alz11a}
M. Alzamora, J. Munevar, E. Baggio-Saitovitch, S. L. Bud'ko, Ni Ni, P. C. Canfield, and D. R. S\'anchez, J. Phys.: Cond. Mat. 23 (2011) 145701.

\bibitem{can16a}
Paul C. Canfield, Tai Kong, Udhara S. Kaluarachchi, and Na Hyun Jo,  Philos. Mag. 96 (2016) 84.

\bibitem{kon15a}
Tai Kong, Sergey L. Bud'ko, and Paul C. Canfield,  Phys. Rev. B 91 (2015) 020507.

\bibitem{ter10a}
Kunihiro Kihou, Taku Saito, Shigeyuki Ishida, Masamichi Nakajima, Yasuhide Tomioka, Hideto Fukazawa, Yoh Kohori, Toshimitsu Ito, Shin-ichi Uchida, Akira Iyo, Chul-Ho Lee, and Hiroshi Eisaki, J. Phys. Soc. Jpn. 79 (2010) 124713.

\bibitem{liu13a}
Yong Liu, M. A. Tanatar, V. G. Kogan, Hyunsoo Kim, T. A. Lograsso, and R. Prozorov, Phys. Rev. B 87 (2013) 134513.

\bibitem{kle16a}
Z. Klencz\'ar, MossWinn 4.0 Manual (2016).

\bibitem{pre12a}
C. Prescher, C. McCammon and L. Dubrovinsky, J. Appl. Cryst. 45 (2012) 329.

\bibitem{gut11a}
Philipp G\"utlich, Eckhard Bill, and Alfred X. Trautwein, {\it M\"ossbauer Spectroscopy and Transition Metal Chemistry. Fundamentals and Applications}, Springer-Verlag, Berlin, Heidelberg, (2011).

\bibitem{via83a}
R. Vianden, Hyperfine Int. 15/16 (1983) 189.

\bibitem{nis76a}
K. Nishiyama, F. Dimmling, Th. Kornrumpf, and D. Riegel, Phys. Rev. Lett. 37 (1976) 357.

\bibitem{ver83a}
H. C. Verma and G. N. Rao, Hyperfine Int. 15/16 (1983) 207.

\bibitem{lon99a} 
Gary J. Long, Dimitri Hutot, Fernande Grandjean, Donald T. Morelli, and Gregory P. Meisner, Phys. Rev. B 60 (1999) 7410.

\bibitem{tam12a}
Ichiro Tamura, Tsuyoshi Ikeno, Toshio Mizushima, and Yosikazu Isikawa, J. Phys. Soc. Jpn. 81 (2012) 074703.

\bibitem{lin11a}
J. Lind\'en, J.-P. Lib\"ack, M. Karppinen, E.-L. Rautama, H. Yamauchi, Solid State Commun. 151 (2011), 130.

\bibitem{che95a}
V.M. Cherepanov, M.A. Chuev, E. Yu. Tsymbal, Ch. Sauer, W. Zinn, S.A. Ivanov, V.V. Zhurov, Solid State Commun. 93 (1995) 921.

\bibitem{boh16a}
A. E. B\"ohmer, A. Sapkota, A. Kreyssig, S. L. Bud'ko, G. Drachuck, S. M. Saunders, A. I. Goldman, P. C. Canfield, preprint,  arXiv:1612.07341 (2016), Phys. Rev. Lett. - in press.

\bibitem{ran11a}
 S. Ran, S. L. Bud'ko, D. K. Pratt, A. Kreyssig, M. G. Kim, M. J. Kramer, D. H. Ryan, W. N. Rowan-Weetaluktuk, Y. Furukawa, B. Roy, A. I. Goldman, and P. C. Canfield, Phys. Rev. B 83 (2011) 144517.

\bibitem{nin08a}
N. Ni, S. Nandi, A. Kreyssig, A. I. Goldman, E. D. Mun, S. L. Bud'ko, and P. C. Canfield, Phys. Rev. B 78 (2008) 014523.

\bibitem{gol08a}
A. I. Goldman, D. N. Argyriou, B. Ouladdiaf, T. Chatterji, A. Kreyssig, S. Nandi, N. Ni, S. L. Bud'ko, P. C. Canfield, and R. J. McQueeney, Phys. Rev. B 78 (2008) 100506.


\end{thebibliography}
\end{document}